\documentclass[%
 reprint,
 amsmath,amssymb,
 aps,
 prl
]{revtex4-2}[prl]

\usepackage{graphicx}
\usepackage{dcolumn}
\usepackage{bm}
\usepackage{placeins}
\usepackage{booktabs}
\usepackage{scalerel}

\DeclareRobustCommand{\emphbf}[1]{\hstretch{0.89}{\textbf{#1}}}

\begin{document}

\title[Physics-Preserving Neural-Network Accelerated Simulations of Plasma Turbulence]{Physics-Preserving AI-Accelerated Simulations of Plasma Turbulence}

\author{Robin Greif}
 \affiliation{Max Planck Institute for Plasma Physics, Boltzmannstr.~2, D-85748 Garching, Germany}

\author{Nils Thuerey}
 \affiliation{Technical University of Munich, Boltzmannstr.~3, D-85748 Garching, Germany}

\author{Frank Jenko}
 \affiliation{Max Planck Institute for Plasma Physics, Boltzmannstr.~2, D-85748 Garching, Germany}

\date{\today}

\begin{abstract}
    Turbulence in fluids, gases, and plasmas remains an open problem of both practical and fundamental importance. Its irreducible complexity usually cannot be tackled computationally in a brute-force style. Here, we combine Large Eddy Simulation (LES) techniques with Machine Learning (ML) to retain only the largest dynamics explicitly, while small-scale dynamics are described by an ML-based sub-grid-scale model. Applying this novel approach to self-driven plasma turbulence allows us to remove large parts of the inertial range, reducing the computational effort by about three orders of magnitude, while retaining the statistical physical properties of the turbulent system.
\end{abstract}


\maketitle

{\it Introduction.} The advent of computing, and now machine learning (ML), has enabled scientists to address challenging scientific questions that have previously been intractable.
One of the most prominent of these is the study of turbulence, with applications ranging from quantum physics~\cite{quantum_turbulence} to astrophysics~\cite{Chamani2017} -- involving liquids, gases, and plasmas~\cite{doi:10.1146/annurev-fluid-010719-060214,perspective_on_turbulence_learning}.
However, even exascale computing will not allow us to tackle some of the most pressing open issues in a brute-force style. 

As it turns out, the combination of computing and machine learning offers some unique opportunities along these lines.
This is the topic of the present Letter.

One popular approach for efficiently computing the dynamics of turbulent systems for a wide range of applications is the Large Eddy Simulation (LES) technique. 
Here, the system is simulated with only the largest scales resolved explicitly, while the unresolved scales are accounted for by a Sub-Grid-Scale (SGS) model~\cite{nature_perspective_cfd_ml, kochkov_machine_2021, les_sgs_mhd}. 
In the following, we will give this old idea a new twist. 
Specifically, we will develop an SGS model based on a Neural Network (NN) with Learned Corrections (LC) on the resolved scales to create a hybrid numerical and ML approach.
As will be demonstrated below, by using a non-propagated field, this approach can be remarkably effective allowing us to cut off virtually the entire inertial range, just retaining the drive range~\cite{turbulence_book}.
This is fundamentally different from previous studies, which focused on the much simpler problem of removing diffusion-dominated scales in the dissipation range~\cite{um2021solverintheloop,kochkov_machine_2021}.
However, removing (large) parts of the inertial range while retaining the integrity of the cascade dynamics has been the major challenge facing LES approaches~\cite{pnas_extreme_events_turbulence}.
Simply extending approaches that work within the dissipation range to the inertial range is typically not a viable option.
In this Letter, we introduce a model that is able to overcome these difficulties and do so very efficiently. 
In fact, it is able to produce physically indistinguishable results even when removing (large) parts of the inertial range, while allowing for a relative speedup of about three orders of magnitude.

{\it Turbulent system.} To illustrate the power of our approach, we will address an open issue of utmost practical importance in the area of contemporary turbulence research -- namely the need to understand, predict, and control turbulent flows that are observed in magnetic confinement fusion plasmas~\cite{stabilizing_turbulence_stellarators,Choi2021}.
To create burning (i.e., self-heated and electricity-producing) plasmas, the energy confinement time of a fusion device needs to exceed a threshold set by the so-called Lawson criterion~\cite{plasma_physics_1}. 
This very quantity is determined by plasma turbulence~\cite{plasma_physics_1,plasma_physics_2}. 
The underlying nonlinear plasma dynamics can be described, e.g., by 
the two-fluid Hasegawa-Wakatani (HW) model~\cite{camargo,grillix}, which
produces quasi-stationary turbulent states, 
whose statistical properties have been studied and documented thoroughly~\cite{camargo}.

The HW model describes the turbulent transport perpendicular to the confining toroidal magnetic field, providing the time evolution of the plasma density $n(x,y,t)$ and the vorticity $\Omega(x,y,t)$:
\begin{eqnarray}
    \partial_t n &=& c_1 \left( n - \phi \right)
                     - \left[ \phi, n \right]
                     - \kappa_n \partial_y \phi
                     - \nu \nabla^{2N} n \,,
            \label{eqn:hw_density} \\
    \partial_t \Omega &=& c_1 \left( n - \phi \right)
                                      - \left[ \phi, \Omega \right]
                                      - \nu \nabla^{2N} \Omega \,.
            \label{eqn:hw_vorticity}
\end{eqnarray}
Here, the Poisson bracket is defined as $[a,b]=\partial_x a\,\partial_y b-\partial_y a\,\partial_x b$, 
and the vorticity $\Omega$ relates to the electrostatic potential $\phi$ via $\Omega = \nabla^2 \phi$. 
The model contains three parameters: the drive strength $\kappa_n$, which describes the degree of inhomogeneity of the background plasma density (in the $x$ direction), the so-called adiabadicity parameter $c_1$, which is inversely proportional to the resistivity of the plasma, and the hyperviscosity parameter $\nu$, which determines the onset of dissipation on small scales.
In the hydrodynamic limit of large resistivity ($c_1\rightarrow0$), the model reduces to the Navier-Stokes (NS) equation. 
This property, in a sense, makes it a more generalizable testing ground that allows to build on top of recent developments in simulating NS turbulence using traditional solvers with learned components~\cite{kochkov_machine_2021,um2021solverintheloop,jax-md}.

\begin{figure}[hbt]
    \centering
    \includegraphics[width=\linewidth]{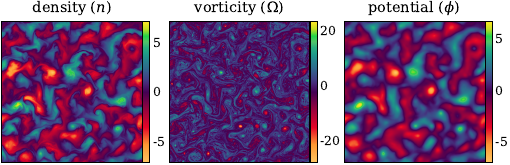}
    \caption{
        Snapshot of a DNS showing the characteristic plasma turbulence of the HW model in the fully saturated turbulent regime (t=300). 
        Note, only $n$ and $\Omega$ are fields propagated in the model and the potential $\phi$ is derived from these.
    }
    \label{fig:snapshot_turb}
\end{figure}

A typical snapshot -- taken in the fully developed turbulent phase of a 2D HW simulation (HW2D) -- 

is shown in Fig.~\ref{fig:snapshot_turb}. 
The observed highly irregular flow patterns in space and time are characteristic of turbulence.
However, the resulting quasi-stationary states far from thermodynamic equilibrium are remarkably robust in a statistical sense~\cite{turbulence}.
For a high-dimensional complex system like this, validation in a strictly deterministic and microscopic sense is meaningless --- any small numerical variation will lead to drastically differently looking states on a very short time horizon.
For studies of plasma turbulence via the HW model, the single most important quantity of interest is the 
turbulent particle flux $\Gamma_n$, which can be written in real-space and spectral coordinates as
\begin{equation}
    \Gamma_n(t) = -\!\! \iint{\! \mathrm{d}^2\! x \;\, n \,\partial_y \phi  }  
                = -\!\! \int{\!\mathrm{d} k_y \; i k_y \, n\scriptstyle(k_y) \; \displaystyle\phi^*\scriptstyle(k_y)\displaystyle } 
    \label{eq:gamma_n}
\end{equation}
where the spectral representation employs averages in $x$ and $\displaystyle\phi\scriptstyle(k_y)\displaystyle^*$ marks the complex conjugate of $\phi$.
During the quasi-stationary turbulent phase, the value of $\Gamma_n(t)$ will fluctuate temporally around a stable long-term mean in a characteristic manner, making it a statistical property of the system.
Therefore, we will employ a statistical paradigm to validate that machine learning imitates the physical effects rather than learning specific states explicitly.
In this approach, an idealized simulator does not only retain visual dynamics and average values, but in fact the underlying distribution from which the physical values are sampled. 
This marks the strongest possible verification of the physicality for complex systems that can be envisioned without analytic solutions and therefore significant stricter evaluation than previously employed.

{\it Novel ML-based LES technique.} Direct Numerical Simulations (DNS) of plasma turbulence tend to require significant computational resources. 
In Cartesian space, for a square box resolving wavenumbers of multiples of $k_0=0.15$, the minimum spatial resolution to retain key physical properties in the HW system is \texttt{512x512}, as careful convergence tests have revealed. The associated time step is $\Delta t = 0.025$, using a 4th-order Runge-Kutta (RK4) scheme.

The hyperviscosity-coefficient $\nu$ of order $N$ and resolution determine the onset of diffusion in the system.
The dissipative scales (which are dominated by diffusion) provide a trivial application space for convolutional neural networks, since diffusion can be expressed via convolutional kernels~\cite{kochkov_machine_2021, um2021solverintheloop}.

Previous LC-SGS approaches have stayed strictly in the diffusion dominated range~\cite{kochkov_machine_2021}, since the approaches breaks down when cutting into the inertial range.
Even here, small differences accumulate through the positive-feedback loops from repeated applications of NNs and had to be mitigated through temporally unrolling on the order of 100 simulation steps~\cite{um2021solverintheloop}.
The approach presented in this Letter, however, is able to remove significant portions of the inertial range, while maintaining the physicality of the system with just three steps unrolled.

\begin{figure}[h!]
    \centering
    \includegraphics[width=\linewidth]{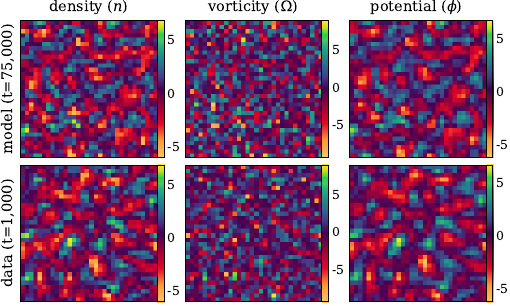}
    \caption{
        Top: PSC Model run continuously for about 3,000,000 steps after being trained on a window of 3 frames. Bottom: Downsampled ground truth DNS at t=1,000. 
    }
    \label{fig:long_run}
\end{figure}

We resolve this fundamental limitation faced by LC-SGS models by introducing the potential-based surrogate correction (PSC). 
By restricting the input to exclude model-propagated fields and projecting the information into the potential instead, 
we constrain the information flow from forming the common positive-feedback loop~\cite{kochkov_machine_2021,um2021solverintheloop}.
This means, the prediction of the low-resolution DNS is transformed into the potential $\tilde{\phi}$ that is fed as a surrogate into the SGS-NN to derive the LC on the model fields $n$ and $\Omega$.
With this approach, we are able to not only keep the simulations of the model stable, but are able to demonstrate that the network performs almost perfectly as an idealized simulator without a fully resolved inertial range.
A sample evaluation visualizing this stability is visualized in the first row of Fig.~\ref{fig:long_run} after running for three million steps.
By allowing us to cut into the inertial range, we are able to reduce the size of the simulation by a factor of 256 from \texttt{512x512} $\rightarrow$ \texttt{32x32}.\footnote{Comparisons for the dissipative range are given in the supplementary information}
We further significantly reduce the computational effort by going from a RK4 scheme to an Euler prediction --- while retaining the timestep size.

This timestep is five times larger than what is required for predictor-corrector schemes, and can be increased by another factor of five without affecting the results. 
However, this property is a phenomena previously encountered~\cite{kochkov_machine_2021,um2021solverintheloop} and as an engineering optimization left for the supplementary information to focus here on the more critical, novel physicality of the solution with under-resolved inertial range. 
As will be shown next, the proposed approach retains the statistical properties of the turbulent system -- which is a remarkable achievement.

{\it Preserving the statistical properties.} We compare different methods in the drastically reduced physical space of \texttt{32x32}, primarily based on their preservation of the critically important turbulent particle flux over time $\Gamma_n(t)$ (see Eq.~\ref{eq:gamma_n}).
In the following, \emphbf{Downsampled} refers to reduced representations of the high-resolution DNS that are provided as ground-truths for training. 
This value is compared to previous approaches that include and learn on the model-propagated fields ($n$, $\Omega$) in the network, 
named \emphbf{previous()}, where the brackets denote the fields used and the number of timesteps unrolled for training.
In our experiments of more than 5,000 trained networks, only about 1\% of these evaluated setups produced stable models.
Therefore, we steel-man our argument by cherry-picking the best of these previous approaches as the baseline \emphbf{previous($n$,$\Omega$,15)}, with two others shown for illustration purposes of the characteristic divergence when extrapolating long-term (\emphbf{previous($n$,$\Omega$,$\phi$,5)}, \emphbf{previous($n$,$\Omega$,5)}).
In contrast, the PSC approach produced stable simulation in every single of the 500+ initialization we attempted even with as few as 3 timesteps unrolled.
We evaluated these for up to $10^6$ times larger timeframes (see Fig.~\ref{fig:long_run}) than they were trained upon.
Additionally, the PSC approach demonstrated no signs of over-fitting even when continuing training for a factor of 100 beyond what produced stable simulations, indicating further robustness of the approach.
Finally, fine-tuned low-resolution DNS are given as references denoted with the fourth-order time integration scheme used \emphbf{DNS(rk4)}.

\begin{figure}
    \centering
    \includegraphics[width=\linewidth]{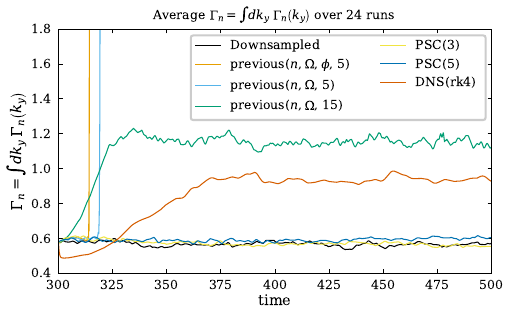}
    \caption{
        Comparison of $\Gamma_n(t)$ for \texttt{32x32} slices of: DNS(rk4) \emphbf{Downsampled} from \texttt{512x512}, a fine-tuned \emphbf{DNS(rk4)}, best versions of \emphbf{previous} LC approaches using different fields and temporal unrolling, and the \emphbf{PSC} approach parametrized in its temporal unrolling. The only previous model that kept stable provided unphysical results by a factor 2, while all other trained versions diverged. 
    }
    \label{fig:stability}
\end{figure}

The timetraces of $\Gamma_n(t)$ are shown in Fig.~\ref{fig:stability}. 
The PSC approach, even with 3 steps of unrolling \emphbf{PSC(3)} shows a mean deviation of less than $0.1\%$ w.r.t.~the reference data \emphbf{Downsampled}. 
Even the cherry-picked best models from previous approaches accumulate errors before reaching an equilibrium, resulting in significantly larger values for $\Gamma_n(t)$. 
The example visualized here found an equilibrium at twice the target value.
In relative terms, therefore, the mean value of PSC simulations are 2,000 times more accurate than the best mean value of \emphbf{previous} models and 700 times more accurate than artificially fine-tuned \emphbf{DNS(rk4)} simulations. 
Our approach accomplishes this feat with only a single gradient computation, compared to four previously, for a theoretically reduction in computational effort by a factor of $10^4$. Empirically, we observe roughly half of this ideal value without introducing specific optimizations. 

\begin{table}
    \centering
    \caption{{Physical values at} \texttt{32x32}}
    \label{table:result_overview}
    \begin{tabular}{@{}lrrrr@{}}
        \toprule
        Variant             & $\Gamma_n\pm\delta\Gamma_n$ & $\Gamma_c\pm\delta\Gamma_c$ & $E\pm\delta E$   & $U\pm\delta U$   \\
        \midrule
        Downsampled        & 0.57$\pm$0.05               & 0.45$\pm$0.03               &  3.69$\pm$0.29  &  8.16$\pm$0.41  \\
        \midrule
        \emphbf{Our Model} & \emphbf{0.57$\pm$0.05}      & \emphbf{0.46$\pm$0.03}      &  \emphbf{3.33$\pm$0.24}  &  \emphbf{8.08$\pm$0.40}  \\
        Previous           & 0.84$\pm$0.10               & 1.11$\pm$0.14               &  8.38$\pm$0.81  & 17.23$\pm$1.61 \\
        DNS(rk4)           & 0.92$\pm$0.13               & 0.51$\pm$0.08               & 10.89$\pm$1.65  & 15.87$\pm$2.20  \\
        \bottomrule
    \end{tabular}
    \footnotetext{Comparison of the physical values for the \emphbf{Downsampled} DNS from \texttt{512x512} to \texttt{32x32} to \emphbf{Our Model} of the PSC, \emphbf{Previous} approaches using the model fields themselves, and fine-tuned \emphbf{DNS(rk4)} all run at \texttt{32x32}: $\Gamma_n$ (particle flux), $\Gamma_c$ (primary sink), $E$ (Energy), and $U$ (Enstrophy).}
\end{table}

This marks a significant improvement to previous methods in two fields: generalizability and efficiency.

First, previous non-LC approaches often relied on explicitly enforcing retention of conservation laws~\cite{pinn,pinn_conservation_laws,prl_network_laws,yu_pinn_turbulent_flow} for systems to remain stable and physical when extrapolating far beyond the time-scales it was trained upon or into non-dissipative regions. 
Our approach contains no model-specific knowledge.

Secondly, generalized LC approaches previously required the network to be exposed to $10^2$ steps that it generated during training~\cite{kochkov_machine_2021,um2021solverintheloop}. 
This vastly increases resource usage for training, thereby limiting the complexity of models it can be applied to. 
This means, previously more than 99\% of data had to be affected by the network, whereas the PSC approach required just over half, or 2 out of 3 time steps in training, to be affected by it. 
Going to fewer steps, e.g., 2, would mean the network is no longer dominantly trained on the data it affected. 
Therefore, the PSC approach reached the lower bound for temporal unrolling, where \textit{just} over half the data generated in a training step is affected by the network. 
At the same time, the PSC data presented here extrapolates orders of magnitude further into the future (far beyond the training time window) than previous studies to demonstrate stability. 
Furthermore, the large windows for training that were previously used introduce implicit time-smoothing that eliminates effects present on temporal scales smaller than the training window. 
Just as a reminder, the PSC approach achieved this while removing significant parts of the inertial range and not simply modeling diffusion-dominated regions, marking a huge step forward towards efficient, physically consistent data-driven methods that currently stunt the widespread adoption in computational sciences.

Moving to a first- and second moment statistical evaluation, table~\ref{table:result_overview} shows physical properties evaluated for 24 simulations over time well within the turbulent phase ($t\in(300, 1000)$), with the mean and standard deviation across 24 simulations separated by $\pm$. 
All PSC values fall well within a single standard deviation of the downsampled DNS, with only the energy being slightly outside one sigma bounds.

This marks the first indication that the PSC-LC are mimicking the contributions of unresolved scales.

In the following, we will solidify this conclusion via additional careful analysis of the hybrid simulations and investigate how far we get towards an idealized generator that preserves all effects of the unresolved physics.

Moving beyond the analysis of scalar quantities, we will now consider the properties of various spectral distributions. 
Obviously, this step represents a significant refinement of our systematic comparison of the direct and hybrid simulations.

\begin{figure}[hbt!]
    \centering
    \includegraphics[width=\linewidth]{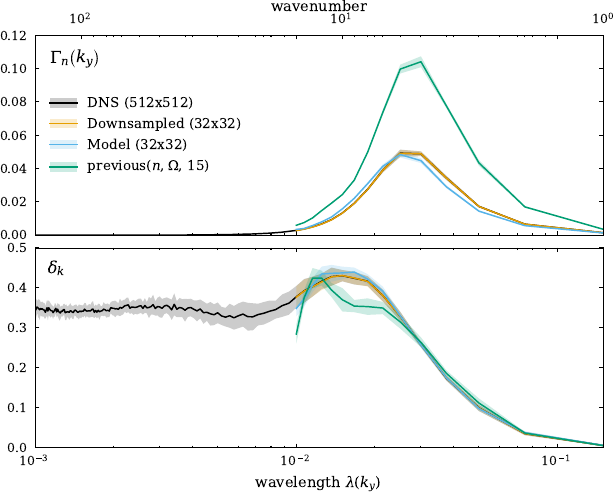}
    \caption{
        The spectral comparison shown is between high-resolution DNS, the downsampled version, our model, and a previous reference model respectively.
        Shaded areas mark $\pm1\sigma$ across 25 simulations of their mean over time.
        Additional spectra are given on the right. 
        Significant improvements in preserving accuracy, precision, and shape can be observed in comparison to previous models across all four spectra.
        Upper: Spectral distribution of $\Gamma_n(k_y)$, describing the strength of the coupling of that wavenumber/wavelength to the background gradient. 
        Lower: drift-wave angle $\delta_k$.
    }
    \label{fig:gamma_n_k}
\end{figure}

Considering once again the turbulent particle flux $\Gamma_n$, it is of great interest to inspect the spectrum $k_y$ to determine the contributions stemming from different spatial scales: $\Gamma_n \!\left(k_y\right) = i k_y\, n \scriptstyle \left(k_y\right) \displaystyle \, \phi^*\! \scriptstyle \left(k_y\right) \displaystyle$ (see Eq.~\ref{eq:gamma_n}). 
Visualizing this wavelength spectrum representation in Fig.~\ref{fig:gamma_n_k}, we expect a distinct bell-shape around $2\times10^{-2}$ from the high-resolution DNS, here called \emphbf{Numerical Integrator (512x512)}. 
Due to coarser resolution, the high-$k_y$ (low wavelengths) tail is not resolved, leading intrinsically to a slight underestimation for $\Gamma_n$ at very low resolution, like in the downsampled case here. 
In these representations, \emphbf{DNS(rk4)} on \texttt{32x32} had to be left out, as the scales would make the images unreadable. 
Previous approaches overestimated the turbulent source across the spectrum to retain stability. 
The PSC, meanwhile, results in a slight shift towards lower frequencies to account for the unresolved turbulent flux contributions of unresolved scales. 
Given the coarse spectral representation, these are still \emph{very} close to the ground truth, especially with respect to the spectral coupling strength at the wavelengths. 

While these changes remain within reasonable limits for $32\times 32$ hybrid simulations, it is clear that a further reduction to, say, $16\times 16$ grid points is not possible -- unless other simulation parameters (like the box size) are adapted. 
Such attempts shall not be pursued here.

Another spectral value that is important in studies based on the HW model, is the phase shift between the two complex-valued variables $n(k_y)$ and $\phi(k_y)$ originating from the linear drift waves, as defined by
\begin{align}
    \delta(k_y) = {\Im}[\log{ n(k_y)^{*} \phi(k_y) }]    
\end{align}
In HW turbulence, like in many other turbulent systems characterized by nonlinearly coupled waves, one tends to find that these distributions are centered around the respective linear values and have a small to moderate width. The $\delta(k_y)$ spectra from our hybrid model and the corresponding direct simulation are plotted in Fig.~\ref{fig:gamma_n_k} (top right). Once again, the model traces the ground truth closely within about one standard deviation throughout the retained scale range of the hybrid simulation.
Previous approaches had significant challenges to retain this shape, resulting in non consistent phase angles, thereby changing the physical dynamics of the system.

Given the phase angles dependence on the model fields in Fourier space, their wavespectra are additionally given in Fig.~\ref{fig:gamma_n_k}.
In this case, they provide additional support that the PSC approach retains the memory of the initial linear phases of the system consistent with the physical model.
It should be noted here that no information about the spectral properties were introduced into the training or design of the system: the network produced the dynamics that preserve spectral properties simply from seeing 3 slices in Euclidean space at a time.
As such, these spectral results provide further evidence towards the physicality of the corrections that the neural network provided, while also visualizing some sources for the discrepancies arising in previous approaches.

\FloatBarrier

Finally, we now show that our hybrid model actually preserves the underlying statistical distributions themselves. 
For turbulent systems, as with all complex systems, the statistical distribution of physical values describes the fundamental physical process underpinning it~\cite{turbulence_velocity_distribution,interstellar_turbulence_distrobution}.
Given the chaotic nature of turbulent systems, no exact value or state can be recreated in finite precision computations.
Therefore, showing that these very distributions are reproduced provides the strongest possible verification of the hybrid model, suggesting that it actually accounts for the relevant physical effects instead of simply creating similar looking data.

To achieve this, we will consider the discrete Cumulative Distribution Function (CDF) of the values of $\Gamma_n$ over time and simulations. 
The discrete CDF is preferred to remove any suggestion of hiding results behind smoothing effects from transforming discrete simulations into continuous probability distribution functions. 
For the values of $\Gamma_n$, this is defined as $F_{\Gamma_n}(x) = P(\Gamma_n\leq x)$ where x is the value of $\Gamma_n$, visualized on the x-axis. 
Therefore, it shows on the y-axis the fraction of values smaller than or equal to the corresponding value on the x-axis, with y ranging from 0 to 1 (100\% of the data is smaller than this value).

\begin{figure}[tb!]
    \centering
    \includegraphics[width=\linewidth]{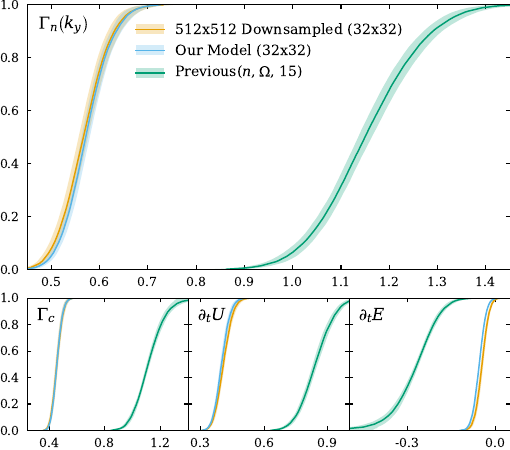}
    \caption{
        Showing the fraction of in values smaller than the value on the x-axes over time (discrete cumulative distribution function)  for the downsampled, our model, and previous approaches, with the integrated $\Gamma_n$ on the top, and $\Gamma_c$, $\partial_t U$, and $\partial_t E$ below from left to right. 
        Once again, means are a solid line with standard deviations across simulations shaded, indicating clearly that not only mean and standard deviation are preserved, but also higher order statistical moments of the distributions from the shape.
    }
    \label{fig:cdf_gamma_nk}
\end{figure}

Considering one last time $\Gamma_n$, Fig.~\ref{fig:cdf_gamma_nk} shows on the top that the cumulative distribution of values recovers the shape within the shaded $\pm$ standard deviation of the ground truth, with almost identical bounds across the entire range of values.
Even when comparing higher-order moments, like the skewness~(0.24~vs~0.22) or the kurtosis~(0.13~vs~0.12), the results from our model and the downsampled high-resolution DNS data are once again very close.
In fact, once adjusted for a small constant offset, statistical tests like the Kolmogorov-Smirnov test cannot reject the null hypothesis that the data were sampled from the same distribution.
Meanwhile, the significant deviation in shape and position of the previous models for the turbulent particle flux is self-evident.

Lastly, we include three other metrics in all figures to further emphasize that the $\Gamma_n$ evaluation metric was not cherry-picked, but that the PSC approach is in fact able to retain the distribution of physical properties that are determined by the dynamics of the turbulent system over time. 
Adopting a source-sink perspective, $\Gamma_n$ takes the role of the turbulent source term in the system. 
The most important of the counteracting sink, $\Gamma_c$, shown on the bottom left in Fig.~\ref{fig:cdf_gamma_nk} exhibits remarkably similar results including its higher order statistical moments- 
To summarize the dynamics, the time variation of the energy $\partial_t E$ and enstrophy $\partial_t U$ of the open system are given to show their similarity. 

In fact, these results for the CDF are representative of the other source and sink terms that define the motion of the system.
These results mark the strongest possible indication that the PSC-based LC model presented here does in fact reproduce the physical effects of unresolved scales even when removing significant parts of the inertial range for the HW model.

{\it Summary.} In this Letter, we have presented, applied, tested, and discussed a potential based NN-SGS method for LES that preserves the physical effects of unresolved scales even when cutting into the inertial range of the turbulent motion.

Our method allows the neural network to efficiently learn how to correct corresponding simulations with a grid 256 times coarser and up to timesteps 5 times larger than was required to produce a converged reference solution.
Meanwhile, it provides results for key physical quantities of the turbulent system that are almost indistinguishable from their DNS counterparts -- not simply visually, in averages, standard deviations, spectrally, but even in terms of their statistical distributions.
These findings provide a significant step forward in the area of turbulence simulations and highlight the value of further developing AI-based techniques for the acceleration of important computational problems.

The approach presented here used as little model-specific information as possible, to strengthen the case for potential generalization to other physical systems. 
This implies, in turn, that many further optimizations for this specific model were potentially left on the table, such as, e.g., the inclusion of information from the frequency domain and known conservation laws. 
In addition, our approach has not yet been optimized for using the least amount of data or targeting the shortest training time.
Despite that, we find an empirical increase of the computational performance by about three orders of magnitude. Further enhancements are left for future work.

\FloatBarrier

\appendix

\nocite{*}

\bibliography{paper}

\appendix

\section{Methods used}

The DNS is implemented in line with previous works~\cite{camargo,grillix} for periodic boundary conditions in all spatial directions.
In each step, the electrostatic potential $\phi$ is reconstructed from the propagated $\Omega = \nabla^2 \phi$ by solving the Poisson equation in frequency space. The indeterminacy in the solution is resolved through setting the offset to zero, given the model definition as perturbation fields.
The HW equation gradient is discretized to second-order accuracy: spatial derivatives are calculated through central finite differences schemes, with the Poisson brackets solved via the Arakawa scheme~\cite{arakawa_computational_1966} to preserve the quadratic invariants (enstrophy, or square vorticity, and kinetic energy, or square velocity).

The baseline DNS simulations were performed with an explicit RK4 scheme in time, with the other only serving as references to compare the hybrid predictor-corrector analogous LC approach presented here. 
The Euler step that was used with the PSC approach required a 5 times smaller timestep to remain stable. Hence, the PSC provided significant stability improvements.

All obtained results were verified through comparisons to previously published research wherever possible.

The numerical implementation of the research-grade framework PhiFlow used for building the code were verified through continuous integration tests for all numerical implementations against native NumPy reference code~\cite{numpy}.

The stability of the HW implementation has been empirically validated and compared through physical metrics to previous reference works~\cite{camargo,grillix}. 

E.g., the stability of the physical property $\Gamma_n$ with resolution to the numerical stabilization coefficient $\nu$ has been visualized in Fig.~\ref{fig:gamma_n_stability} with more plateaus visualized in Fig.~\ref{fig:plateaus}. These clearly show that the hyperviscosity coefficient does not affect the physical values for resolutions at \texttt{512x512} or above, for $\nu$ between $(10^{-6},10^{-5})$ at that resolution acrosss all physical values. 
For further details, see Supporting Information.

In total, over a petabyte of reference data was created through this process to establish verifiable stable references with dependable uncertainties across the parameter space.
Given the large amount of resources, we plan on making the data publicly available as a reference dataset for other researchers to train and test future work.
For the DNS LES to converge on its own, a minimum resolution of 512x512 was required (see Appendix~\ref{subsec:appendix_verification}) using fourth-order temporal discretization.  
While the high-resolution is necessary to preserve the dynamic contributions in time, each state can be compressed through downsampling in frequency space and still preserve the physical properties relatively well. 
Specifically, key contributions to $\Gamma_n$ stem from relatively large wavelengths such that at \texttt{32x32} less than one standard deviation of $\Gamma_n$ is lost (see Appendix~\ref{subsec:appendix_downsampling}). 
The coarser \texttt{32x32} grid therefore represents the domain for LES in this work, for which the corrections are learned.
Downsampling this drastically from \texttt{512x512} to \texttt{32x32} cuts into the inertial range of the turbulent flow. 
Meaning, the effects of the no longer numerically resolved wavelengths and their physical effects on the dynamics have to be accounted for by the PSC network in the hybrid AI LES approach to create stable physical results. 
Higher resolutions, cutting less into the inertial range like \texttt{64x64}, trivially preserve these properties better and are included as comparison in the Supporting Information, but are less strong of a case for the methodology than the more demanding case of \texttt{32x32} presented here.

\begin{figure}[t!]
    \centering
    \includegraphics[width=\linewidth]{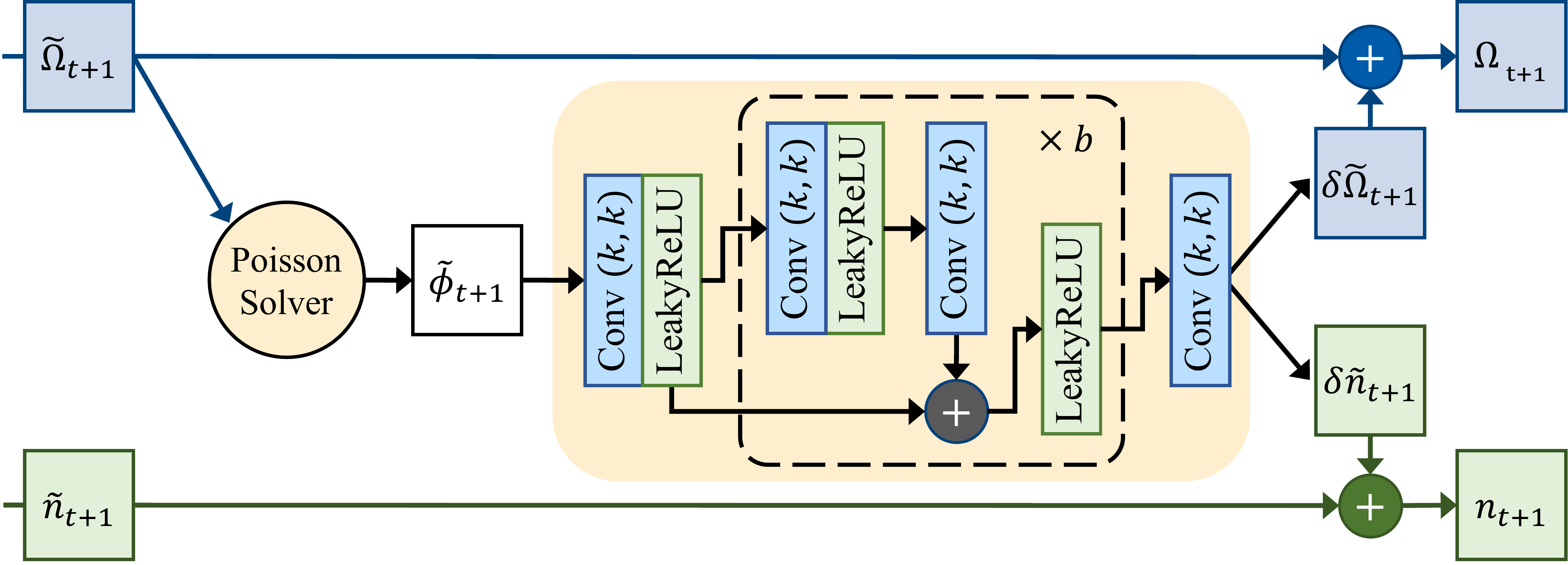}
    \caption{
        Information flow for the hybrid AI LES scheme focusing the PSC network corrector. Specifically, the potential $\phi_t$ is calculated from the vorticity $\Omega_t$ through the Poisson solver. The gradients $\partial n_t$ and $\partial \Omega_t$ at time $t$ are calculated through finite differences. Progression with an Euler step, gives the predictions $\tilde{n}_{t+1}$ and $\tilde{\Omega}_{t+1}$ shown on the left. The potential-based surrogate corrector (PSC) feeds only the predicted potential $\tilde{\phi}_{t+1}$ into the network, which returns the corrections $\delta \tilde{n}_{t+1}$ and $\delta \tilde{\Omega}_{t+1}$ that are then added to the prediction to generate the next state.
    }
    \label{fig:psc_architecture}
\end{figure}

The sub-grid system consists of the choice of input and output fields, network architecture, and their composition - with the standard LC presented using an Euler-step that is analogues to the adding a correction ($\delta \tilde{\Omega}_{t+1}$, $\delta \tilde{n}_{t+1}$) to the DNS prediction ($\tilde{\Omega}_{t+1}$, $\tilde{n}_{t+1}$). 
This is illustrated in Fig.~\ref{fig:psc_architecture}.

The PSC improves on previous LC models by not including any of the propagated fields (density and vorticity) in the model, and opting for including the potential of the system instead as the input.
The potential provides an intuitive description of the system that contains sufficient information for the network to generate consistent corrections of the model propagated fields as its the output.
By using only the potential as the input to correct a system that does not propagate the potential, but only the density and vorticity, we created a significant layer of abstraction into what has previously been a straight forward positive-feedback loop. 

This change in the information flow practically eliminated feedback within the system, thereby negating a main drawback when using neural networks within simulations that otherwise has to be mitigated.

The designed information flow was stable across variations in the network architecture and composition. 
Different architectures were tested for the network itself, but with this twist the established Residual Network (ResNet) architecture provided sufficient accuracy, precision, and hyperparameter stability to not warrant further search (see appendix~\ref{sec:appendix_neural_network}).
Specifically, the presented architecture is using two dimensional convolutions (Conv2D) kernels of size $(5,5)$, with $32$ filters each, and $b=5$ central blocks with a leaky rectified linear unit (LeakyReLU) activation.
A single block of the architecture is visualized inside Fig.~\ref{fig:psc_architecture}.
The network was trained using an Adam optimizer with a learning rate of $10^{-5}$. 
These reference values give sufficient stability for initialization, however, variations likewise yield comparable results.

The learning approach used in this study builds on the work of temporally unrolling multiple steps during training that has demonstrated more physically consistent results for simulations of classical fluids. However, these improvements come at the cost of requiring significantly more resources for training~\cite{um2021solverintheloop}. 

Time integrated learning unrolls individual steps of the hybrid predictor-corrector as a chain to predict multiple time steps into the future before comparing to ground truth. 

This method exposes the network to the data that it generates thereby helping to reduce feedback.
As a result, the model can generate physically consistent data for much longer time spans than those it is trained upon~\cite{um2021solverintheloop}.
In previous setups this approach required extremely long chains of model timesteps in memory that make it computationally extremely expensive to train and --- for extremely complex turbulence simulations as found in fusion research --- unfeasible.
The goal of time integrated learning is to expose the network to the effects it has on the results.
This means that to train it dominantly on its own effects, there exists a lower bound where just over half of the generated time steps fed into the network were corrected by it.  
Using the PSC approach, we were able to achieve
stable simulations at this lower-limit for LC methods, with as little as 3 consecutive time steps unrolled (see Fig.~\ref{fig:long_run}), marking orders of magnitude reduction in computational effort compared to previous studies~\cite{um2021solverintheloop}. 
All results shown in the paper are achieved at this lower bound.

\section{Hasegawa-Wakatani}

\subsection{Two-dimensional Hasegawa-Wakatani equations}
\label{subsec:HW}

The Hasegawa Wakatani (HW) equations constitute a two-fluid model for drift wave turbulence in magnetized plasmas. Restricting the set of differential equations to the two dimensions perpendicular to the background magnetic field, one arrives at Eqs.~\ref{eqn:hw_density} and \ref{eqn:hw_vorticity} in the main text.

In our  simulations, the drive strength $\kappa$ was kept fixed at unity, facilitating the verification of our results w.r.t.~published results~\cite{camargo}. A key model parameter is the adiabatic coefficient $c_1$, characterizing the coupling strength between the plasma density and the electrostatic potential by means of resistive effects. In the hydrodynamic limit, $c_1\rightarrow0$, the HW model reduces to the two-dimensional Navier-Stokes equations, which were investigated in a previous study~\cite{um2021solverintheloop}. In the adiabatic limit, $c_1\gg1$, one obtains the so-called Hasegawa-Mima equation~\cite{hasegawa_mima} (which is isomorphic to the Charney equation in geophysical fluid dynamics), with weaker turbulence.

Therefore, the cross-over region between these two limits with values of $c_1\sim 1$ is of particular interest, and has also been the subject of previous studies~\cite{grillix,camargo}.

To avoid the accumulation of fluctuation energy on the smallest spatial scales, numerical dissipation is applied via a hyperviscosity term. 
The latter is parameterized via the diffusion coefficient $\nu$ and an order $N$, both of which have to be hand-tuned depending on the choice of the physical parameters, numerical resolution, and applied numerical methods: big enough to suppress accumulation, small enough not to affect the quantitative results.
As it turns out, at sufficiently high resolution, there exists a finite region in $\nu$ space, for which the results do not depend on this parameter. However, this region shrinks with decreasing resolution and ultimately disappears. This effect defines the minimum resolution for a given set of physical parameters.
While different studies proposed different parametrizations, such as $\nu_\phi \neq \nu_n$ and $N_\phi \neq N_n$, these modifications merely shift the plateau region by some amount. In line with much of the existing literature, we employed $\nu = \nu_\phi = \nu_n$ and $N=3$. 

A grid search was performed using a few thousand test runs with variations on these that have verified the apriori argument presented here.
For the baseline low resolution DNS --- called DNS (\texttt{32x32}) --- the hyperviscosity was fine-tuned to give a value close to the downsampled version. As it can be seen in Fig.~\ref{fig:stability} and Fig.~\ref{fig:plateaus}, these do exist, but since they are dominated by the viscosity do not provide physically convergent results and are purely artificially kept numerically stable.

\begin{figure}[tbhp!]
    \centering
    \includegraphics[width=\linewidth]{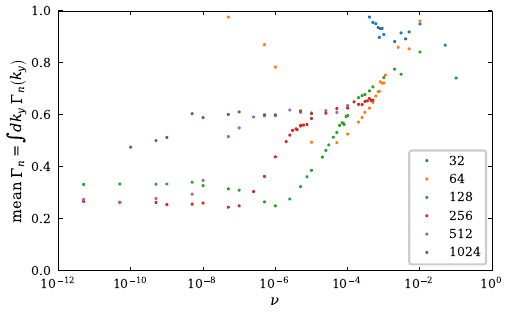}
    \caption{
        Values for $\Gamma_n=\int{\!\mathrm{d}k_y \Gamma_n(k_y)}$ integrated in Fourier space on the y-axis, over varying numerical stabilization strengths on the x-axis, for a number of resolutions (colored samples) keeping $c_1=1$ and $k_0=0.15$ fixed. 
        It shows resolutions of 512x512 with hyperviscosity parameter values between $10^{-5}-10^{-6}$ are stable against variation of the latter.
    }
    \label{fig:gamma_n_stability}
\end{figure}

\subsection{Verification}
\label{subsec:appendix_verification}

\subsubsection{Reference data generation}
\label{subsubsec:data_generation}

The reference data for the present work was generated via the numerical implementation described in Section~\ref{subsec:numerical_implementation}.

The initialization was sampled from a random normal distribution for each grid point, with amplitudes scaled down by dividing values by 100 [$\mathcal{N}(\mu=0, \sigma=10^{-4})$].

Since only explicit time integration schemes were used, the temporal discretization was derived empirically for each set of parameters using grid search, starting from the largest and decreasing continuously until a large stable step size was encountered.

For hyperviscosity, we used $N=3$, corresponding to 6th-order gradients, following established references~\cite{camargo}.
As discussed above, for each set of physical parameters, a $\nu$ scan was carried out, ensuring that the resulting physical values are not dependent on this choice.

The grid search for stable parameters to ensure the simulations have converged to physical stable values can be seen for the value of $\Gamma_n$ in  Fig.~\ref{fig:gamma_n_stability}. 
In the figure it can clearly be seen that for a $c_1=1$ and a physical box size of $k_0=0.15$, a resolution of $512\times 512$ was needed, with a time step of $dt=0.025$ with an RK4 scheme and stable for a hyper-diffusion parameter $\nu$ between $10^{-6}$ and $10^{-5}$. 
For comparison, the predictor-corrector time integration schemes that are the numeric analogue to the LC approach required a 5$\times$ smaller timestep to be stable.

For this set of parameters, the linear growth phase transitions into the early turbulent phase around $t=75$.
A quasi-steady state with fully developed turbulence is established around $t=200$. The training data was taken from the time window t=300-1000, implying that each simulation required 40,000 time steps at the required $dt$.
Previous studies~\cite{grillix, camargo} stopped at shorter end times. This meant that although the results from these did provide results in the correct ballpark, as was sufficient for those studies, the values were not necessarily converged to their physical average and hence a later end-time was chosen here. Even then, drift of the mean could still be observed in some runs, therefore multiple runs had to be taken into consideration to account for variable convergence times.
These reference data generation required a vast amount of data to be generated. For example, a single $512\times 512$ dataset at $c_1=1$ using single precision generated 42 GB for each of the three fields to store as reference data. This obviously drastically increased for higher resolutions and shorter time steps required.
In order to estimate the standard deviation across initializations, each then had to run once again multiple times with different initializations to ensure no conflating variables were introduced.

\subsubsection{Physical verification of DNS implementation}
\label{subsec:appendix_physical_verification}

To facilitate comparisons of our high-resolution DNS with reference literature and learning analysis, we will introduce some important metrics for the HW model. Describing an open physical system, energy is constantly injected into the system, nonlinearly redistributed among the available degrees of freedom, and dissipated. In a saturated turbulent state, a flux equilibrium is reached, in which the average injection and dissipation rates balance each other.

The energy $E$ of the system and its time derivative are given by

\begin{align}
    E &= \frac{1}{2} \iint{\! \mathrm{d^2} x \; 
         \left(n^2 + \left| \nabla \phi \right|^2 \right)}
         \label{eqn:energy} \\
    \frac{\partial E}{\partial t} &= \Gamma_n - \Gamma_c - \mathfrak{D}^E
                                         \label{eqn:e_dt}
\end{align}

Meanwhile, the enstrophy $U$ and its time derivative are given by

\begin{align}
    U &= \frac{1}{2} \iint{\! \mathrm{d}^2 x \; 
         \left(n - \Omega \right)^2} 
         \label{eqn:enstrophy} \\
    \frac{\partial U}{\partial t} &= \Gamma_n - \mathfrak{D}^U 
                                     \label{eqn:u_dt}
\end{align}

Here, the source and sink terms are defined as

\begin{align}
    \Gamma_c &=  c_1 \iint{\! \mathrm{d}^2 x \; 
         \left(n - \phi\right)^2} 
    \label{eqn:gamma_c} \\
    \mathfrak{D}^E &= \quad \iint{\! d^2 x \; (n\, \mathfrak{D^n} - \phi\, \mathfrak{D}^\phi)} 
    \label{eqn:DE} \\
    \mathfrak{D}^U &= - \iint{\! d^2 x \; (n - \Omega)(\mathfrak{D}^n - \mathfrak{D}^\phi)} 
    \label{eqn:DU}
\end{align}

Given various differences in the numerical implementation and grid resolution, the values of these quantities in the saturated turbulent state found in literature vary within certain limits (see Table~\ref{table:references}).

\begin{figure*}[h!]
\centering
    \includegraphics[width=0.495\linewidth]{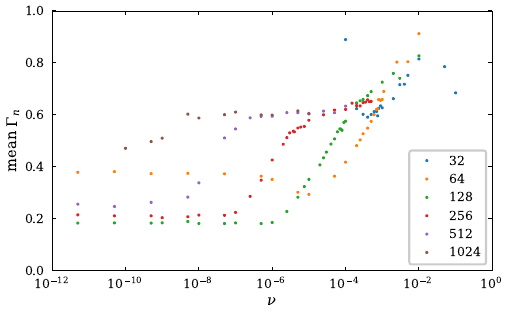}\hfil
    \includegraphics[width=0.50\linewidth]{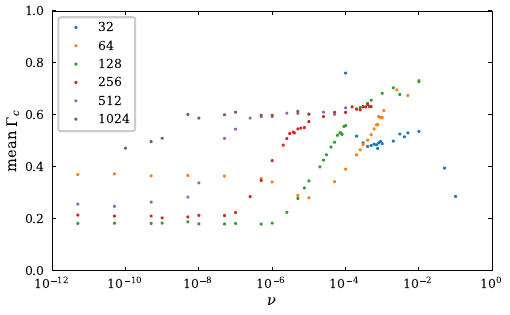}\par\medskip
    \includegraphics[width=0.50\linewidth]{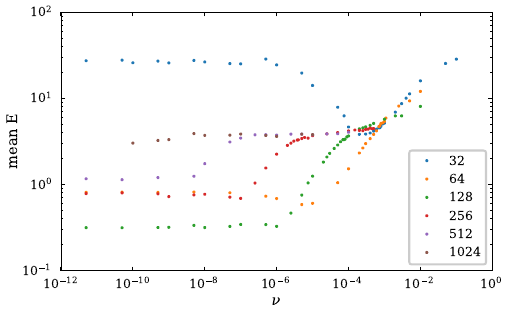}\hfil
    \includegraphics[width=0.50\linewidth]{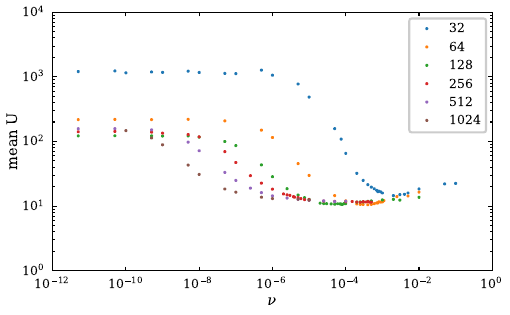}
    \caption{
        Mean values of $\Gamma_n$ (top left), $\Gamma_c$ (top right), $E$ (bottom left), and $U$ (bottom right) for resolutions ranging from \texttt{32x32} to \texttt{1024x1024} with varying values for the hyperviscosity coefficient $\nu$ for fixed $c_1=1$ and $k0=0.15$. On the linear scale in the first row it becomes clear that \texttt{512x512} is the first resolution where a true plateau is established such that $\nu$ has no effects on the physical value for at least a full order of magnitude.
    }
    \label{fig:plateaus}
\end{figure*}

In the context of the present study, great care was taken to ensure convergence and quantify uncertainties, so to provide robust and reliable results.
The obtained values are listed in Table~\ref{table:references}. 
Here, we averaged over time $t=300$ until $t=1,000$ and calculated the mean and standard deviation across 24 simulations. We selected the \texttt{512x512} simulations with $\nu\in(10^{-6},5\times10^{-6},10^{-5})$.
The stability of these simulations can be verified for the entire set of physical values presented using Fig.~\ref{fig:plateaus}, where a clear plateau for these values is visible.
The values in Table~\ref{table:references} show clearly that the physical values established using our high-resolution DNS agree very well with the most recent study of Stegmeir\cite{grillix}, and are in the ballpark of Camargo\cite{camargo} and HW~\cite{grillix}.

\begin{table}[tb!]
    \centering
    \caption{
        Physical metrics: mean values $\pm$ standard deviation
    }
    \begin{tabular}{@{}lrrrrr@{}}
        Metric     & Our Data         & Stegmeir        & Camargo        & HW              & Zeiler  \\
                   & (\texttt{512x512}) & \cite{grillix}  & \cite{camargo} & \cite{grillix}  & \cite{zeiler}  \\
        \midrule
        $\Gamma_n$ & $0.60 \pm 0.01$  & $0.64$                  & $0.73$                & $0.61$            & $0.8$                \\
        $\Gamma_c$ & $0.60 \pm 0.01$  & $n/a$                   & $0.72$                & $n/a$             & $n/a$                \\
        $E$        & $3.78 \pm 0.07$  & $3.97$                  & $4.4$                 & $3.82$            & $6.1$                \\
        $\delta E$ & $0.29 \pm 0.03$  & $0.26$                  & $0.16$                & $0.26$            & $0.51$               \\
        $U$        & $13.2 \pm 0.91$  & $n/a$                   & $12.8$                & $n/a$             & $n/a$                \\
        $\delta U$ & $0.54 \pm 0.08$  & $n/a$                   & $1.66$                & $n/a$             & $n/a$                \\
        \bottomrule
    \end{tabular}
    \label{table:references}
\end{table}

In this work, we often refer to $\Gamma_n$ as the integral of $\Gamma_n(k_y)$ in $k_y$ space, as opposed to the integral of $\Gamma_n(y)$ in $y$ space. While these two quantities are always identical for our model, they can differ in the DNS case at low resolution. Therefore, we chose to use  $\int \mathrm{d} k_y \Gamma_n(k_y)$ for comparisons between the model results and their DNS counterparts.

\subsubsection{Validating downsampling methods}
\label{subsec:appendix_downsampling}
\begin{figure}[h!]
    \centering
    \includegraphics[width=\linewidth]{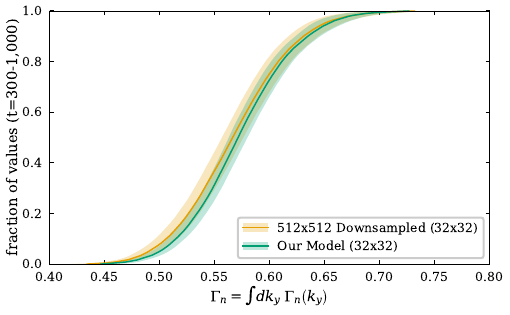}
    \caption{The spectrally integrated electron flux $\Gamma_n=\int{\!\mathrm{d}k_y \Gamma_n(k_y)}$ across downsampling factor along each axis over time. In the legend the average and standard deviation is given $t\in(300, 1000)$. 
    }
    \label{fig:fourier_downsampling}
\end{figure}

The high-resolution data generated on a \texttt{512x512} grid was mapped onto a coarser grid using downsampling in two-dimensional Fourier space. 
This implies that we are cutting off the turbulent energy cascade, as can be seen in spectral analysis (i.e., Fig.~\ref{fig:gamma_n_k}).
Empirically, downsampling in Fourier space provides advantages over mean, median, and skipwise downsampling in terms of its effects on key observables of the system under investigation.

Overall, the data shown in Fig.~\ref{fig:fourier_downsampling} confirms the expected behavior:  $\Gamma_n$ decreases monotonically with increasing downsampling, since certain spectral components that contribute to the turbulent source are removed.
Specificaly, downsampling along each axis in Fourier space by a factor of 8, for a compression of 64, leads to an integrated $\Gamma_{n}$ only marginally smaller than in the high-resolution DNS simulation at $<\!1.7\%$.
Furthermore, using a factor of 16, for a compression factor of 256, allows to have deviation from the high-resolution DNS values for the integrated $\Gamma_{n}(k_y)$ within the bounds of a single standard deviation.

The downsampling also retains the half max autocorrelation time of the density and potential field within statistical uncertainty with a mean of $t_\mu=2.6$ and standard deviation $t_\sigma=0.16$ for the density, and $t_\mu=2.8$ and standard deviation $t_\sigma=0.18$ for the potential.

Finally and trivially, this form of downsampling preserves all spectral components of the original datasets fields in the best possible way. As a result, the spectra of various derived quantities of interest are also retained.

\begin{figure*}[tb!]
    \centering
    \includegraphics[width=\linewidth]{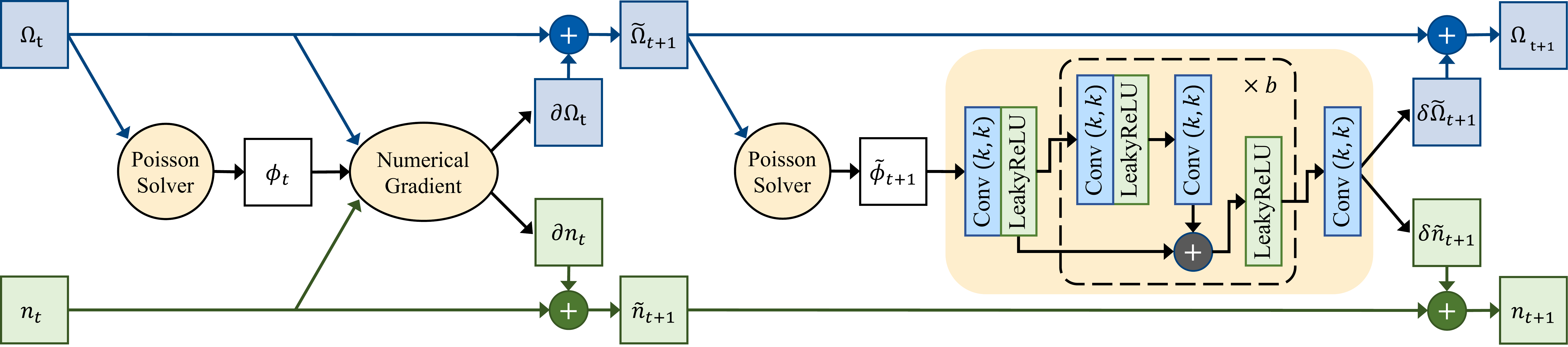}
    \caption{Diagram of information flow through the numerical predictor and neural corrector to generate a full step of the PSC based LES scheme.}
    \label{fig:full_architecture}
\end{figure*}

\section{Neural Network}
\label{sec:appendix_neural_network}

\subsection{Network Implementations}
\label{subsec:appendix_network_implementation}

The network architectures selected for the corrector were all based on convolutional operators, due to their direct translation to numerical recipes currently employed in state-of-the-art plasma simulations and their previous successful demonstration for classical fluids~\cite{um2021solverintheloop}.

The network architectures were all implemented using TensorFlow 2.x~\cite{tensorflow2015-whitepaper}, including updated versions of networks discussed in Ref.~\cite{um2021solverintheloop}.

Since TensorFlow does not support periodic boundary conditions for convolution kernels at the writing of this paper, a custom layer was implemented that created an explicit circular padding in a differentiable manner before every convolution.

\subsubsection{ResNet}
\label{subsubsec:appendix_resnet}

The primary network architecture deployed in the present work is a convolutional neural network called the Residual Network (ResNet) that has been found to perform exceptionally well in previous similar applications~\cite{um2021solverintheloop,kochkov_machine_2021} and general imaging problems~\cite{resnet}. 
The simplicity and similarity to standard numerical methods highlighted the potential applicability for the issue at hand.
Recently, it has been shown that ResNets are able to perform at similar levels than more complex modern architectures like the transformer~\cite{resnet_vs_transformer}.
The implementation was parametrized for the number of filters $f$, the number of blocks of layers $b$, and the filter kernel size $k$ to perform the hyper-parameter search on the network architecture.
At the time of developing the code, Tensorflow convolutional layers did not support periodic boundary conditions, meaning that custom layers were implemented to allow for differentiable explicit padding before each convolutional kernel is applied.
This can introduce performance penalties that are not present in other frameworks supporting periodic boundary conditions.
Since the primary objective was to show the viability and rough time speedup, a complete rewrite of the learning implementation was deemed unnecessary.
For a thorough performance analysis we recommend using a framework that supports periodic boundary conditions out of the box.

\subsubsection{UNet}
\label{subsubsec:appendix_unet}

The ResNet implementation was compared against a standard Tensorflow U-Net Implementation that have shown exceptional results in among other things Biomedical Image Segmentation~\cite{unet} and previous LES turbulent flow simulations~\cite{yu_pinn_turbulent_flow}.
The U-Net architecture is defined by reducing the size of latent space representations iteratively, until a pre-determined minimal size is reached and then expanded again, while latent space layers with similar sizes are connected via skip connections. The resultant shape when sketched on a diagram looks like the namesake U-shape, with the down and upscaling going on the perimeter and the skip connections horizontally.
The performance of various U-Net variations was not superior to the simpler ResNet configurations. As a result, although used for similar tasks, they have not been able to reproduce the precision and stability of the ResNet implementation. Therefore, these results have been left out of the main paper, but mentioned as insight for future development.

\subsection{Flow of information through the hybrid network}
\label{subsec:information_flow}

In the main text, we discussed the information flow from the prediction through the potential based surrogate corrector network. A complete depiction of the flow of information is given in Fig.~\ref{fig:full_architecture}. The figure shows how the previously given prediction $\tilde{n}_{t+1}$ and $\tilde{\Omega}_{t+1}$ are generated. First, the potential was solved from the initial vorticity. Then all three initial fields are given to the numerical integrator based on finite difference methods to return a gradient. In the predictor step, an Euler step will be performed to produce the predictions that are passed on to the corrector. For DNS, the predictor was kept the same, but the Euler time stepping scheme was replaced by an RK4 to produce the final states directly $n_{t+1}$ and $\Omega_{t+1}$.

\subsubsection{Network Training}

Considering the amount of information contained for training within each simulation, we calculated the autocorrelation phase to estimate how much uncorrelated data is present.
In the turbulent phase, the mean time it takes for the Pearson correlation coefficient of the physical fields to drop by 50\% is about $t=2.5$. This suggests that the duration of each simulation corresponds to several hundred autocorrelation times.
However, for training, a single simulation has not provided sufficient independent samples to generate converged models; for this, three simulations were needed.

\subsubsection{Training data pre-processing}

All input data for the training of the network was normalized using the standard deviation within a frame, averaged across the selected turbulent regime (t=300-1,000) and averaged across all available simulations for each field. 

One should note that these values might differ slightly from the true standard deviation for some setups, since the input to the network is not the ground-truth data itself, but rather the ground-truth data propagated by one step, using a simulator. 
In practice, this presents no concerns, however, given that the difference was found to be less than $0.1\%$ across the training set. The resulting data has been shown to be statistically indistinguishable from the training data for all fields.

\subsection{Further Model Evaluations}

\begin{table*}[htb]
    \centering
    \caption{Physical values: Mean and Standard Deviation}
    \setlength\tabcolsep{4pt}
    \resizebox{\textwidth}{!}{
    \begin{tabular}{@{}clrrrrrrrr@{}}
        Resolution & Variant                   & $\Gamma_n$     & $\delta \Gamma_n$ & $\Gamma_c$     & $\delta \Gamma_c$ & $E$             & $\delta E$       & $U$             & $\delta U$       \\
        \midrule                                                                                   
        512x512    & Numerical Integrator      & 0.60$\pm$0.01  & 0.05$\pm$0.004    & 0.60$\pm$0.01  & 0.042$\pm$0.004 &  3.79$\pm$0.07  & 0.292$\pm$0.032  & 13.17$\pm$0.90  & 0.548$\pm$0.079  \\
        64x64      & Downsampled (512x512)     & 0.60$\pm$0.01  & 0.05$\pm$0.004    & 0.55$\pm$0.01  & 0.040$\pm$0.004 &  3.76$\pm$0.07  & 0.291$\pm$0.032  &  9.63$\pm$0.25  & 0.438$\pm$0.054  \\
        \emphbf{64x64}      & \emphbf{Our Model}                 & \emphbf{0.61$\pm$0.01}  & \emphbf{0.05$\pm$0.004}    & \emphbf{0.54$\pm$0.01}  & \emphbf{0.040$\pm$0.003} &  \emphbf{4.30$\pm$0.07}  & \emphbf{0.320$\pm$0.038}  &  \emphbf{9.47$\pm$0.09}  & \emphbf{0.450$\pm$0.056}  \\
        64x64      & Numerical Integrator      & 0.55$\pm$0.00  & 0.06$\pm$0.005    & 0.36$\pm$0.002 & 0.057$\pm$0.005 &  1.09$\pm$0.01  & 0.449$\pm$0.028  & 19.28$\pm$0.13  & 2.384$\pm$0.182  \\
        32x32      & Downsampled (512x512)     & 0.57$\pm$0.01  & 0.05$\pm$0.004    & 0.45$\pm$0.01  & 0.036$\pm$0.003 &  3.69$\pm$0.29  & 0.046$\pm$0.020  &  8.16$\pm$0.45  & 0.408$\pm$0.037  \\
        \emphbf{32x32}      & \emphbf{Our Model}                 & \emphbf{0.57$\pm$0.01}  & \emphbf{0.05$\pm$0.005}    & \emphbf{0.46$\pm$0.01}  & \emphbf{0.034$\pm$0.003} &  \emphbf{3.33$\pm$0.24}  & \emphbf{0.055$\pm$0.020}  &  \emphbf{8.08$\pm$0.44}  & \emphbf{0.399$\pm$0.034}  \\
        32x32      & Numerical Integrator      & 0.92$\pm$0.01  & 0.13$\pm$0.010    & 0.51$\pm$0.01  & 0.089$\pm$0.007 & 10.89$\pm$0.19  & 1.651$\pm$0.136  & 15.87$\pm$0.21  & 2.197$\pm$0.164  \\
        32x32      & Previous ($n,\Omega$, 15) & 0.84$\pm$0.01  & 0.10$\pm$0.041    & 1.11$\pm$0.01  & 0.149$\pm$0.014 &  8.38$\pm$0.07  & 0.810$\pm$0.043  & 17.23$\pm$0.13  & 1.611$\pm$0.143  \\
        \bottomrule
    \end{tabular}
    }
    \\
    \label{table:full_table}
\end{table*}

\subsubsection{Visual}
\label{subsec:appendix_visual}

A representative snapshot of the kinds of data sets generated by the network is presented in Fig.~\ref{fig:snapshot_turb}. The distributions of the values of the various fields over time match closely.

One noticable exception is a slow drift in the mean value of $\Omega$ in the implemented version of the PSC models, due to the lack of a normalization of the offset bias. However, this effect does not change any physical attributes, as the fields in the HW model are defined as perturbations on top of the background. Hence, any drift can simply be integrated into the background and ignored. For other models, an explicit normalization to remove the constant bias should be introduced.

\subsubsection{Spectral}
\label{subsec:appendix_spectral}

In studies based on the HW model, it is customary to also investigate the phase shift between the two complex-valued variables $n(k_y)$ and $\phi(k_y)$, as defined by
\begin{align}
    \delta(k_y) = {\Im}[\log{ n(k_y)^{*} \phi(k_y) }]    
\end{align}
In practice, samples of this quantity are taken in the saturated turbulent phase as a function of the radial coordinate $x$ and time $t$, and the resulting distribution functions are plotted. In HW turbulence, like in many other turbulent systems characterized by nonlinearly coupled waves, one tends to find that these distributions are centered around the respective linear values and have a small to moderate width. The $\delta(k_y)$ spectra from our hybrid model and the corresponding direct simulation are plotted in Fig.~\ref{fig:gamma_n_k} (top right). Once again, the model traces the ground truth closely within about one standard deviation throughout the retained scale range of the hybrid simulation, 

while even stable previous approaches using the model-fields themselves diverge over half of the spectrum.

\subsubsection{Distributions}
\label{subsec:distributions}

The distributions shown in Fig.~\ref{fig:cdf_gamma_nk} of the main text show clear alignment of our approach with the downsampled dataset. 
While small shifts of the distribution center between the model and downsampled reference can sometimes be observed (e.g., in the case of $\partial_t E$), the shape of the distribution is always preserved. 
Moreover, this effect disappears when using less aggressive downsampling and/or longer temporal unrolling (i.e., using 5 timesteps instead of the showcased 3).
Overall, the consistent preservation of the shapes across all physical metrics of the model is far clearer than previous work in the field, to the point that plotting the reference of previous models visually eliminates any small differences. 

A thorough comparison of the mean of all metrics and their standard deviations across 24 simulations is further presented in Table~\ref{table:result_overview} including the fine tuned DNS. Once again all metrics, and even their standard deviations acrosso runs are preserved.

\end{document}